# Electronic Structure of Strained Manganite Thin Films with Room Temperature Ferromagnetism Investigated by Hard X-ray Photoemission Spectroscopy

Hidekazu Tanaka [1, 4, *], Yasutaka Takata [2], Koji Horiba [2], Munetaka Taguchi [2], Ashish Chainani [2], Shik Shin [2], Daigo Miwa [2], Kenji Tamasaku [2], Yoshinori Nishino [2], Tetsuya Ishikawa [2], Mitsuhiro Awaji [3], Akihisa Takeuchi [3], Tomoji Kawai [1] and Keisuke Kobayashi [3]

[1] *ISIR-Sanken, Osaka University, 8-1 Mihogaoka, Ibaraki, Osaka, 567-0047, Japan*
[2] *RIKEN/SPring-8, 1-1-1, Mikazuki-cho, Sayo-gun, Hyogo 679-5148, Japan*
[3] *JASRI/SPring-8, 1-1-1 Kouto, Mikazuki-cho, Sayo-gun, Hyogo 679-5198, Japan*
[4] *PRESTO, Japan Science and Technology Agency, 4-1-8 Honcho Kawaguchi, Saitama, Japan*

We report the bulk sensitive Hard X-ray ($h\nu$=5.95keV) core level photoemission spectroscopy to investigate the intrinsic electronic structure of strained $(La_{0.85}Ba_{0.15})MnO_3$ thin films. In a 20nm thick well-strained film with strongly enhanced ferromagnetism, a new sharp satellite peak appeared at the low energy site of the Mn $2p_{3/2}$ main peak, whereas a broader signal was observed for the unstrained film with 300nm thickness. Cluster calculations revealed that the intensity corresponded to the density of the state at the Fermi level relating to the magnitude of the ferromagnetic order. The satellite intensity also agreed quantitatively with the square of the magnetization.

---

Perovskite manganite exhibits a rich variety of electric and magnetic properties including colossal magnetoresistance [1], ferromagnetism with metallic conduction [2], perfect spin polarization [3], charge/orbital ordering [4] due to the strong coupling between spin, and charge and orbital (lattice) degrees of freedom [5]. Interestingly, we found that the tensile strain from the substrate stabilized the double exchange ferromagnetism on lightly doped $(La_{1-x}Ba_x)MnO_3$ thin films [6, 7], and that ultra thin films, even those with a 5nm thickness, displayed room temperature ferromagnetism in comparison to a ferromagnetic Curie temperature ($T_C$) of 260K in unstrained systems [8]. This feature is also important in relation to applications pertaining to spintronic devices such as a ferromagnetic $(La,Ba)MnO_3/Pb(Zr,Ti)O_3$ field effect transistor [9] and a ferromagnetic $(La,Ba)MnO_3/Sr(Ti,Nb)O_3$ $p$-$n$ diode [10] working at room temperature. In an effort to elucidate the origin of this enhanced ferromagnetism, information concerning the electronic structures of the films is essential, in addition to that of the lattice structure and macroscopic electric/magnetic properties, and could lead to the development of strongly corrected electron and spin devices [11].

Photoemission spectroscopy (PES) is a powerful and standard tool that has been utilized to investigate the electronic structure of materials such as manganites [12-15]. Nevertheless, the standard PES technique is very surface sensitive even if Soft X-rays (SX) with relatively high energy ($hv$ -- 1keV) are used because of short mean free paths of the emitted low kinetic energy electrons. This technique sometimes yields contradictory results in relation to bulk physical properties. Recently, high-resolution PES studies using Hard X-rays (HX: $hv$ -- 6keV) have been employed in the investigation of $HfO_2$ on Si [16], GaN [17], GaAs [17], (Ga,Mn)N [18], $(V,Cr)_2O_3$ [19], high $T_C$ cuprate [19] and $(La_{1-x}Sr_x)MnO_3$ [20], making it possible to reveal the true bulk electronic structure up to depths of 5nm to 10nm.

We utilized bulk sensitive HX-PES in the investigation of $(La,Ba)MnO_3$ thin films with various film thickness that exhibited strain-enhanced ferromagnetism. The Mn 2$p$ core level HX-PES spectra revealed a systematic change in electronic structure corresponding to the bulk nature of these films. We also proposed a quantitative relationship between the HX-PES spectra and bulk magnetization against the temperature in this system.

(La$_{0.85}$Ba$_{0.15}$)MnO$_3$ epitaxial thin films with thickness ($t$) corresponding to 300nm, 20nm and 3nm were deposited on etched 0.01wt% Nb-doped SrTiO$_3$ (001) single-crystal substrates using a pulsed laser deposition (PLD) technique (ArF excimer: lamda=193 nm) at a substrate temperature of 700°C under an O$_2$ pressure of 0.1Pa. The target employed was a sintered (La$_{0.85}$Ba$_{0.15}$)MnO$_3$ pellet prepared by a conventional solid reaction process. Following film formation, the films were annealed at 750°C for 10 h in an O$_2$ atmosphere of 1 atm. Film structures were examined by X-ray diffraction and high-resolution transmission electron microscopy [7]. Examination by atomic force microscopy (AFM) confirmed that the surface morphology was atomically flat [8]. The magnetic properties were measured using a superconducting quantum interference device (SQUID) magnetometer, and the electrical resistivity was measured using a four-probe method. From Fig. 1, which shows the dependence of magnetization on temperature, it was determined that the $T_C$ of the films were 282K for $t=$ 300nm, 299K for $t=$ 20nm and about 100K for $t=$ 3nm. This tendency depicting an enhancement in ferromagnetism by a strain effect was consistent with our previous reports [6-8]. Following *ex-situ* magnetization measurements, the films were transferred into a photoemission chamber with a base pressure of 10$^{-8}$ Pa from air without any surface cleaning process. HX-PES measurements for the Mn 2$p$ core level were performed at an undulator beamline BL29XU [21-23] of SPring-8 from 28K to 320K. The experimental setup including beamline optics has been described elsewhere [16, 18]. The excitation energy was set to 5.95 keV with a bandwidth of 70 meV. An electron energy analyzer (Gammadata Scienta Co., R4000-10KV) dedicated for HX-PES was used. The total energy resolution was set to 160 meV and was confirmed by fitting the Fermi edge profile of an Au plate. The inelastic mean free path (IMFP) of the Mn $2p$ photoelectrons (Kinetic Energy :-- 5300 eV) in (La,Ba)MnO$_3$ was experimentally estimated as 6 nm. In an effort to verify the validity of the HX excitation for investigations concerning the type of thin films employed in this study, soft x-ray (SX) PES spectra of the Mn $2p$ core level were also measured at an undulator beamline BL17SU. The excitation energy was 1.44 keV and the IMFP was estimated as 1.5 nm.

Figure 2 (a) shows the Mn $2p$ core level spectra of (La,Ba)MnO$_3$ thin films with $t = 300$nm by Hard X-rays with an incident photon energy of 5.65keV (spectrum (A)), $t = 20$nm films by Hard X-rays with an incident photon energy of 5.65keV (spectrum (B)), and $t = 20$nm films by Soft X-ray with an incident photon energy of 1.44keV (spectrum (C)). In the core level spectra of (A) and (B) as measured by HX-PES, additional shoulder (satellite) structures appeared at the low binding energy side (about 639eV) of the Mn $2p_{3/2}$ main peak (about 641eV), in comparison to the SX-PES spectrum (C) at low temperature. This dependence of satellite intensity on temperature was more clearly observed by HX-PES ($h\nu=5.65$keV) than by SX-PES ($h\nu=1.44$keV) for the $t = 20$nm film, and indicated that HX-PES is superior to SX-PES in providing certain information relating to whole film (bulk) properties from an electronic structure standpoint.

Figure 2(b) shows the Mn $2p_{3/2}$ HX-PES spectra at 28K for (La,Ba)MnO$_3$ films with $t= 300$nm, 20nm and 3nm. The satellite intensity of the $t= 20$nm film was much stronger than that of the $t= 300$nm film. This tendency was consistent with the observed enhancement in ferromagnetism of the 20nm film with higher $T_C$ and lower resistivity relative to the 300nm film as shown in Fig. 1. For the $t=3$nm film where the ferromagnetism was strongly suppressed, no shoulder structure was observed. The observed variation in satellite intensity strongly corresponded to the magnitude of the magnetization and metallicity of the (La,Ba)MnO$_3$ films with various thickness. Figure 3 shows the dependence of the Mn 2$p_{2/3}$ spectra on temperature for the $t$ = 20nm film whose ferromagnetism was strongly enhanced. With increasing temperature from 28K to 320K, the intensity of the satellite peak systematically decreased, and almost disappeared at 300K which corresponded to a $T_C$ of 299K for the 20nm film. Obviously, the appearance of the satellite peak corresponded to the evolution of double exchange ferromagnetism in the manganites. K. Horiba *et al* reported that satellite peaks were clearly observed in a (La$_{1-x}$Sr$_x$)MnO$_3$ thin film displaying ferromagnetism (0.2 -- x -- 0.4) but suppressed in antiferromagnetic phases (x=0 and 0.55) in the Mn 2$p$ core level spectra as observed by HX-PES, and concluded that the satellite peak had originated from core-hole screening through evolution of density of the state at the Fermi level ($D(E_F)$) using a MnO$_6$ (3$d^4$) cluster model calculation with D$_{4h}$ symmetry [19]. The observed spectra in Fig. 2 account for the development of $D(E_F)$ relating to double exchange ferromagnetism with decreasing film thickness from 300nm to 20nm. Furthermore, the valence deviation of Mn ion, which could be detected as chemical shift of Mn 2$p_{3/2}$ main peak, is also important to discuss the ferromagnetism. If the excess hole doping, namely the formation of Mn$^{4+}$, is introduced, a shift of the high binding energy side of the main peak toward higher energy would be detected as confirmed in the series of (La$_{1-x}$Sr$_x$)MnO$_3$ films with various Sr$^{2+}$ concentration [20]. A comparison of the observed spectra for the 20nm and 300nm films shown in Fig. 2(b) revealed that the Mn 2$p_{3/2}$ main peaks did not shift with decreasing film thickness and that the satellite intensity was enhanced. This indicated that $D(E_F)$ evolved without additional hole doping. Although the origin of the enhancement of $T_C$ in thinner (La,Ba)MnO$_3$ films has been explained in terms of strain-induced changes in the lattice structure that lead to modification of the $e_g$ band state [6], some reports have argued that the enhancement of $T_C$ originates from oxygen over-doping (namely cation deficiency) that causes additional hole doping [24]. The observed spectra of the 20nm and 300nm films deny an oxygen-over doping scenario, but support a strain scenario. This also agrees with the result that the hole concentration estimated at 10K was almost constant at $6\times10^{20}$ cm$^{-3}$, although the hole mobility increased from 5 cm$^2$/Vs to 50 cm$^2$/Vs with decreasing film thickness from 700nm to 20nm ($T_C$ increased from 260K up to 320K) [25]. Contrary to this, the Mn 2p$_{3/2}$ main peak for the 3nm film, which exhibited strongly suppressed ferromagnetism and insulating behavior, shifted to the higher binding energy side (about 643eV), so that the origin of the suppressed ferromagnetism can be accounted for not by a strain scenario, but by other factors such as a surface boundary layer [3, 26], and charge transfer between the film and substrate [27, 10] considering the shorter IMFP compared to the film thickness.

In an effort to quantitatively clarify the electronic state relating to the spin state from the HX-PES core level spectra, Fig. 4 shows the dependence of the normalized satellite intensity ($I_S(T)$) on temperature obtained by subtraction from the spectrum at 300K. With developing ferromagnetic order and metallicity as the temperature decreased, $I_S(T)$ systematically increased. Cluster calculations which reproduced the observed spectra were also employed to estimate the development of $D(E_F)$ (*C* state was introduced to represent a coherent metallic band at $E_F$, shown in inset of Fig. 3) [19, 20]. Three configurations were used for the initial states, namely $3d^4$, $3d^5\underline{L}$ where $\underline{L}$ is a hole in the ligand O 2*p* states, and $3d^5\underline{C}$ which represents a charge transfer from the *C* state to the Mn 3*d* state. We fit the experimental spectra by changing only two parameters, the charge transfer energy between the Mn 3*d* and the *C* states (delta\*) and the hybridization between the central Mn 3*d* orbitals and the *C* states (*V*\*). Except for these two parameters, all other parameter values were fixed and determined from previous work [20]: the *d-d* Coulomb interaction of the Mn 3*d* states: $U$ = 5.1 eV, the charge transfer energy between the Mn 3*d* and ligand O 2*p* states: delata = 4.5 eV, the hybridization between the Mn 3*d* and ligand O 2*p* states; $V$ = 2.94 eV, the crystal field splitting: $10Dq$ = 1.5 eV, the Coulomb interaction between the Mn 3*d* and Mn 2*p* core hole states: $U_{dc}$ = 5.4 eV. The $V^*$ and delta* parameters obtained are listed in Table I. These could reproduce the intensity and position of the satellite peaks of the experimental Mn 2p HX-PES spectra for the $t$= 20nm film at various temperatures (Fitting curves are not shown. See details in Ref. [20] for the calculation of (La$_{1-x}$Sr$_x$)MnO$_3$ at low temperature.) It can be seen that $V^*$ increased with developing ferromagnetic ordering as the temperature decreased, and indicated that the well-screened feature of $I_S(T)$ was strongly related to the hole doped $D(E_F)$ (**C** state). Analogy with the Kondo coupling between the *f* and conduction band states provides the following relationship [28],

$$V^* \propto \sqrt{D(E_F)} \qquad --- \quad (1)$$

Normalized $(V^*)^2$ plotted against temperature showed a consistency with the experimental behavior of $I_S(T)$, and indicated that $D(E_F)$ was proportional to $I_S(T)$

Our final goal was to estimate bulk (whole film) physical properties, including spin information, from the core level HX-PES spectra. By analogy with the double exchange model described by the mean field approximation [29], $D(E_F)$ is expected to be proportional to the square of the magnetization,

$$D(E_F) \propto (M)^2 \qquad --- \quad (2)$$

As a result, following relationship can be expected.

$$I_S(T) \approx (M)^2 \qquad --- \quad (3)$$

As shown in Fig. 4, the experimental $I_S(T)$-$T$ curve agrees well with the $\{M(T)\}^2$ –$T$ curve as measured by the SQUID magnetometer, so that the HX-PES spectra for the Mn 2p core level could provide electronic structure information corresponding to double exchange ferromagnetism in the bulk part of the (La,Ba)MnO$_3$ film based on eq. (3). In another technique employed to evaluate the spin-related electronic structure for manganite, J.-H. Park *et al* reported on the X-ray absorption magnetic circular dichroism (X-MCD) in addition to surface sensitive spin-resolved PES analysis of a (La,Sr)MnO$_3$ film whose surface was cleaned by an annealing processes in an ultrahigh vacuum chamber [26]. Compared with their X-MCD result concerning the temperature dependency as shown in the inset of Fig. 4, $\sqrt{I_s(T)}$ obtained from the HX-PES spectrum agrees well with the SQUID magnetization value, $M(T)$. This is derived from the bulk sensitive nature of the HX-PES technique. In an effort to check the perfect matching to $M(T)$, employment of a HX-PES experiment using higher photon energy up to 10keV would be interesting.

In conclusion, bulk sensitive core-level HX-PES was used to investigate (La$_{0.85}$Ba$_{0.15}$)MnO$_3$ strained thin films with various thickness. The dependence of the well-screened feature (satellite peak) of the Mn 2$p_{3/2}$ peak on film thickness systematically agreed with the ferromagnetic order in the bulk part of the films, and revealed that $D(E_F)$ of the strained (La,Ba)MnO$_3$ films systematically evolved without additional Mn$^{4+}$ formation in going from the 300nm to 20nm thick films, supporting a strain-induced ferromagnetism scenario from the viewpoint of electronic structure. Furthermore, the satellite intensity observed by HX-PES corresponded to the bulk square of the magnetization of the ferromagnetic (La,Ba)MnO$_3$ thin film on the basis of a model calculation that included charge transfer from the $D(E_F)$ to Mn 3$d$ states.

Perovskite manganite film is considered to be one of the best candidates for use in the development of room temperature spintronics devices such as tunneling magnetoresistance devices [30] and ferromagnetic field effect transistors [9]. For these strongly corrected electron devices, it is essential that the electronic structure of the internal part of the film and interface part of the heterostructure far from the surface be elucidated and understood. Furthermore, insofar as device fabrication is concerned, materials should be processed by post annealing, lithography, electrode deposition and so on. The core-level HX-PES technique allows for an estimation of the bulk electronic structure even in *ex-situ* measurements without any surface treatment. In particular, for ferromagnetic manganite, this technique makes it possible to estimate the electronic structure directly related to the magnetization at deep depths, and we believe that it can play an important role in the development of novel functional magnetic, electrical and optical devices constructed using transition metal oxides.

Table I  Fitting parameters obtained by cluster calculations for the Mn $2p_{3/2}$ HX-PES spectra of the (La,Ba)MnO$_3$. thin film (20nm) at various temperatures.

| Temperature [K] | $V^*$ [eV] | delat*[eV] | Temperature [K] | $V^*$[eV] | delta*[eV] |
|---|---|---|---|---|---|
| 28  | 0.4  | 0.8 | 240 | 0.33 | 0.8 |
| 100 | 0.38 | 0.8 | 260 | 0.32 | 0.8 |
| 160 | 0.37 | 0.8 | 280 | 0.29 | 1.0 |
| 180 | 0.35 | 0.8 | 300 | 0.25 | 1.1 |
| 200 | 0.34 | 0.8 | 320 | 0.24 | 1.2 |
| 220 | 0.34 | 0.8 | | | |

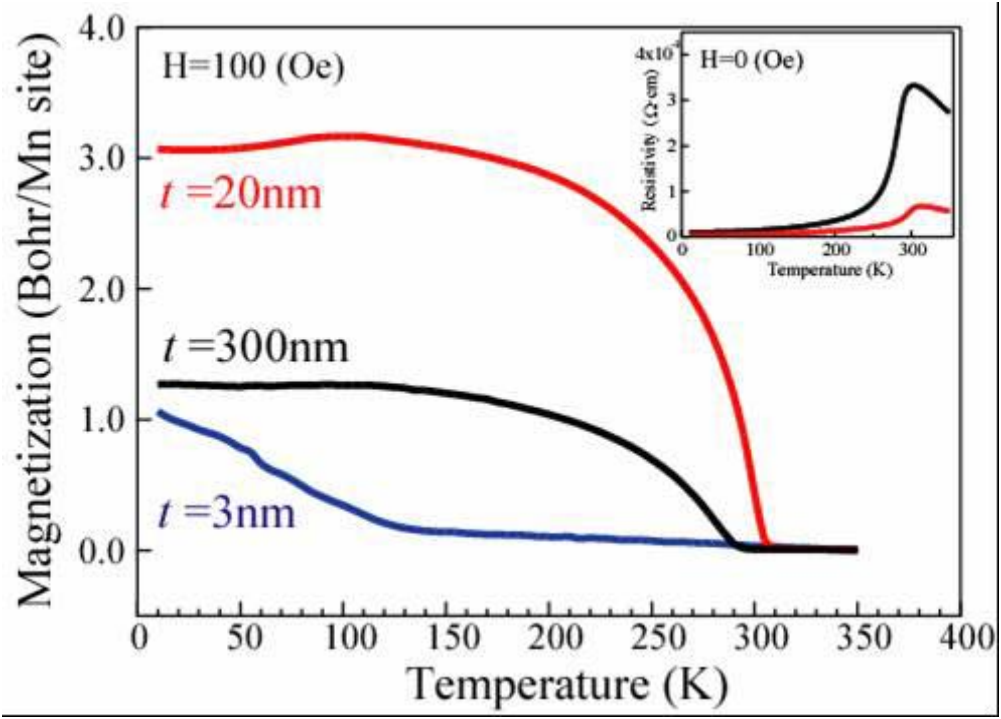

Fig.1 Dependence of magnetization on temperature for $(La_{0.85}Ba_{0.15})MnO_3$ epitaxial thin films with thickness of 300nm, 20nm and 3nm on an Nb-doped $SrTiO_3$ $SrTiO_3$ (001) single crystal substrate under a magnetic filed of 100Oe. $T_C$ was defined as $dM(T)/dT=0$. The inset shows the dependence of resistivity on temperature for the 300nm and 20nm films.

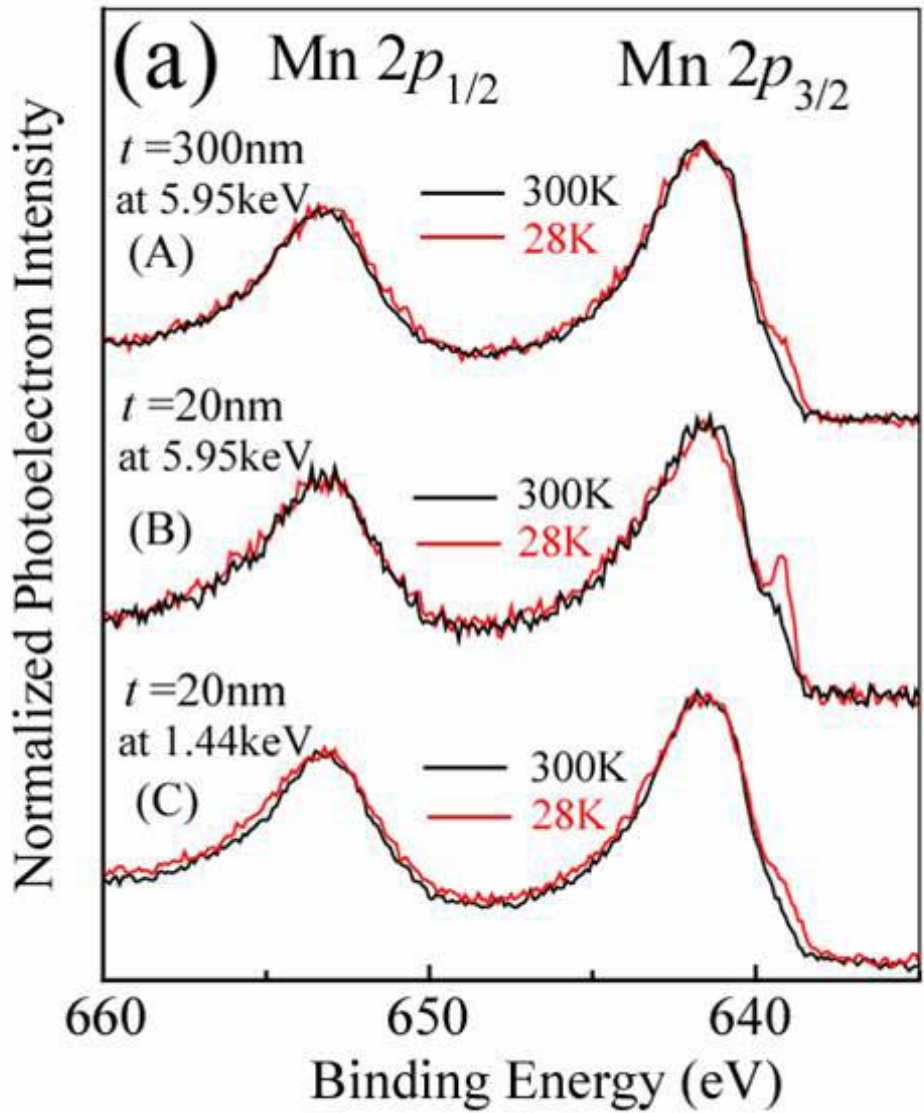

Fig. 2 (a) The Mn 2p core level spectra of $(La_{0.85}Ba_{0.15})MnO_3$ thin films with $t=$ 300nm and $t=$ 20nm by Hard X-ray ($h\nu =$5.95keV), and the spectrum measured by Soft X-ray ($h\nu=$1.44keV) for $t=$ 20nm.

.

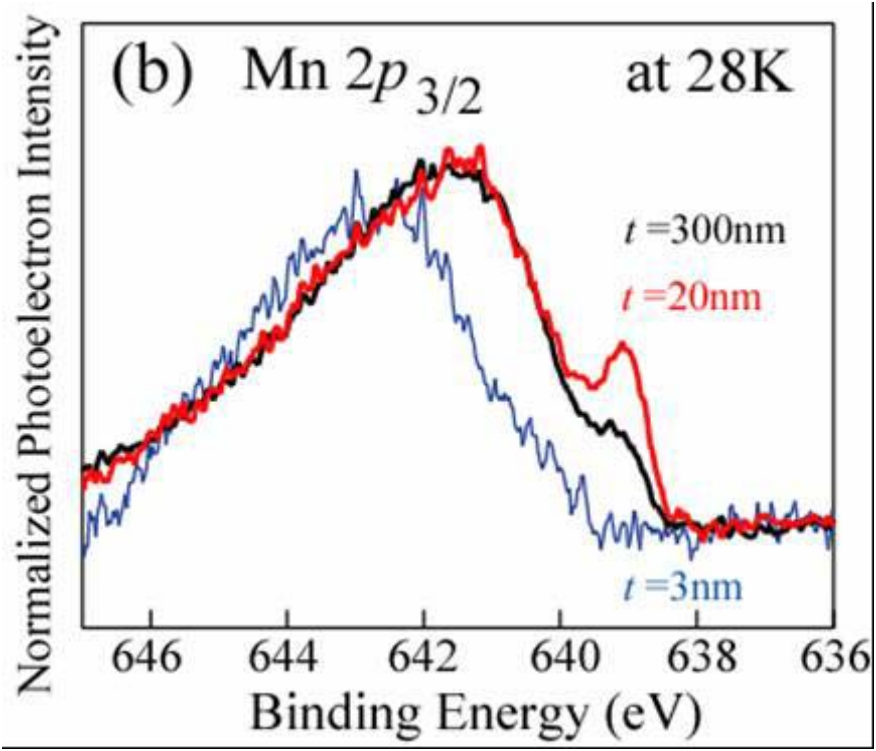

Fig. 2(b) The Mn $2p_{3/2}$ core level spectra of $(La_{0.85}Ba_{0.15})MnO_3$ thin films at 28K with $t=$ 300nm, 20nm and 3nm.

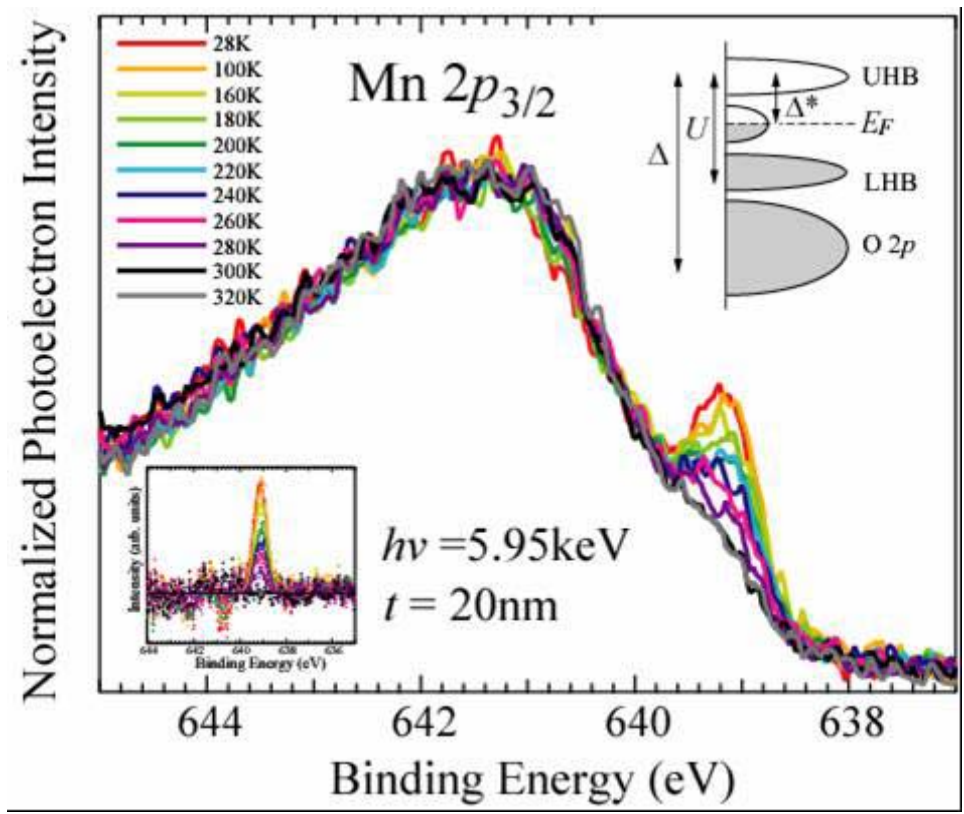

Fig. 3 Dependence of Mn $2p_{3/2}$ spectra on temperature for the $(La_{0.85}Ba_{0.15})MnO_3$ film with 20nm thickness. The insets show the temperature variation part of the satellite peaks subtracted from the spectrum at 300K, and a schematic energy diagram of the valence band.

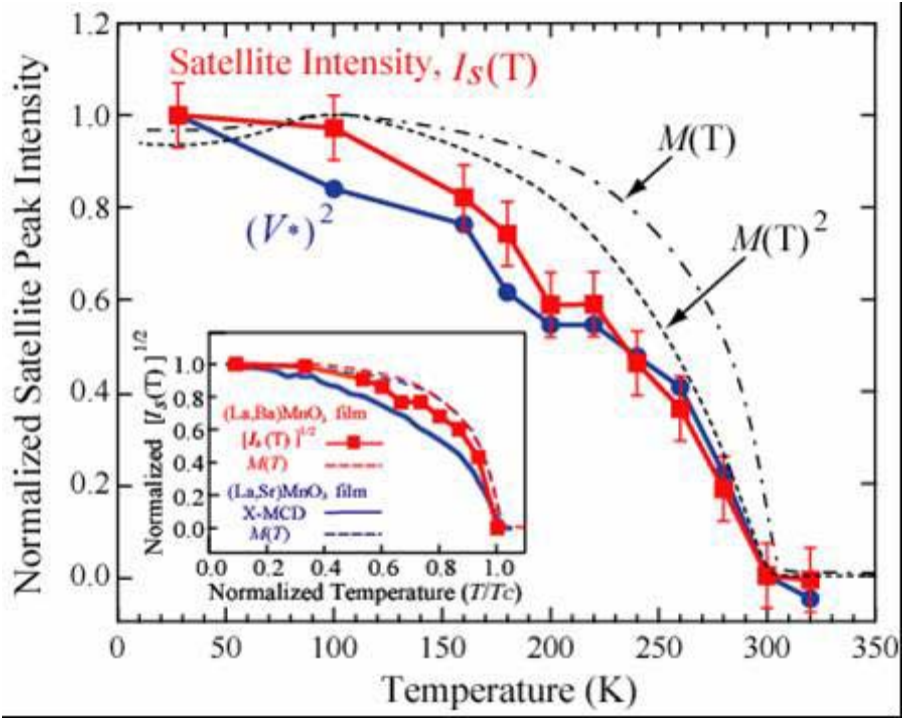

Fig. 4 Summarized dependence of normalized satellite intensities of the Mn $2p_{3/2}$ spectra ($I_S$(T)), normalized hybridization parameter $(V^*)^2$, magnetization ($M$(T) and $M$(T)$^2$) on temperature. $I_S$(T) and $(V^*)^2$ are normalized at 28K (as 1.0) and 300K (as 0). The inset shows the dependence of $\{I_S(T)\}^{1/2}$ and $M$(T) on temperature for the $(La_{0.85}Ba_{0.15})MnO_3$ film ($t$= 20nm), and the dependence of the MCD signal and $M$(T) on temperature for the $(La_{0.3}Sr_{0.3})MnO_3$ film. Data concerning the MCD signal and $M$(T) for the $(La_{0.3}Sr_{0.3})MnO_3$ film were obtained from Ref. [26]